# To Measure In-plane conductivity of Nafion membrane with general electrochemical approach

Jian-Wei Guo[1]*, Jian-Long Wang[1]*   Shang-kun Jiang[2]   Li Li[2]*

1. *Institute of Nuclear and New Energy Technology, Tsinghua University, Beijing, 100084, China*

2. School of Chemistry and Chemical Engineering, Chongqing University, Chongqing 400044, China

\* Corresponding author: E-mail address: jwguo@mail.tsinghua.edu.cn, wangjl@mail.tsinghua.edu.cn,   liliracial@cqu.edu.cn

Tel.: +86 1080194009; Fax: +86 1080194009

**Highlight:**

- The equivalent circuit is key for electrochemical measurements and analysis

- Our equivalent circuit was validated by both EIS and CV measurements.

- Our equivalent circuit has good in situ characterization for measuring system

- The circuit elements revealed kinetic relationship for contact resistance, in-plane conductivity, active and inactive proton in membrane

**Abstract**

It is important to measure in-plane conductivity of Nafion membrane for fuel cell, but this target is generally inhibited by measuring system with heterogeneous interfaces and immature electrochemical measurements. This paper simply used water media to establish stable measuring system with metal electrode and Nafion membrane, representing system as equivalent circuit. Our equivalent circuit was validated by both cyclic voltammetry (CV) and electrochemical impedance spectroscopy (EIS) measurements, also clarified connection and difference in two measurements. This electrochemistry breakthrough helps realize measuring system completely and reliably even under successive temperature cycles, providing circuit elements for kinetic analysis in contact resistance, in-plane conductivity, inactive and active proton. We also clarified that the inactive and active proton shift can dictate low frequency inductance, which is an important sign for active and stable operation in fuel cell, secondary battery and materials. All these results can induce enormous progress in multi-disciplines, making our work have great significance and broad impact for future studies.   .

1. INTRODUCTION

Nafion membrane is a proton-conducting polymer, with main chain of -$CF_2$- segments and side chain of $HSO_3^-$ group. Since the $HSO_3^-$ group can transport proton ($H^+ \cdot H_2O$) in water media, the Nafion membrane is core material in polymer exchange membrane fuel cells and electrolysis cells [1-3]. Therefore, it has great motivation to measure Nafion membrane conductivity, whose proton transport behavior can guide material and cell progress significantly [4-10].

Generally, the Nafion membrane is a 2D (2-Dimension) material with large plane (cm scale) and thin thickness (10-100μm), thus it may have anisotropic conductivity either along plane (in-plane) or along perpendicular direction (through-plane). Both conductivities are important for fuel cell, but it is difficult to measure through-plane conductivity due to measurement sensitivity in thickness direction [11-19]. Even so, it is still important to measure the in-plane conductivity, helping develop large cell and providing characterization frequency for cell operation. Nevertheless, there are many problems to obtain reliable in-plane conductivity for Nafion membrane due to following restrictions. Firstly, the in-plane conductivity is mainly affected by many internal properties of membrane, including membrane thickness, ion exchange capacity, water uptake and channels, gas/liquid permeability and chemical/physical stability [20-23]. Past studied tried to measure conductivity under humidified gas condition, but they usually confront complex water-membrane balance. This operation mode lays a large difficulty to obtain stable in-plane conductivity, other than its kinetic relationship under temperature effect [11,13, 24,25].

Secondly, most studies used metal electrode to establish contact interfaces with membrane, thus composing of measuring system with metal electrode and membrane. Since the metal electrode and membrane transfer electron and proton, respectively, their contact interfaces must experience continuous or discontinuous flowing for heterogeneous charges. This inhibits to obtain stable membrane conductivity, presenting challenge issues in measurement [13, 17, 18, 22, 26, 27].

Thirdly, most studies used electrochemical measurement methods, thus both DC (Direct current) and AC (Alternative current) methods with different approaches can be used. On the one hand, the CV measurement is a typical DC method, but it can only obtain insufficient information, making it as reference measurement [15, 22, 28]. On the other hand, Electrochemical Impedance Spectroscopy (EIS) is an advanced AC (Alternative current) method, and it mainly uses multi frequency to reveal measuring system completely, promising to separate electron and proton process efficiently. Unfortunately, the EIS is still immature in reliable measurement and analysis, and it often dictates complex electric responses with resistances, capacitance and inductance. In most cases, these complex electric responses are fitted with equivalent circuit, but this fitting approach usually has multi equivalent circuits, incapable to achieve effective analysis [5,6,15,28]. Consequently, the EIS is still a large obstacle to obtain reliable in-plane conductivity for Nafion membrane, requiring its core breakthrough in electrochemistry.

To address above issues, this paper simply selects water media as operation condition, which can hydrate Nafion membrane sufficiently and ignore particular properties of membrane. Furthermore, this simple water condition helps membrane and metal electrodes to establish stable measuring system, representing as equivalent circuit. With a core idea in electrochemistry, Scheme 1 proposed that the equivalent circuit dictates various electric responses in CV and EIS measurements, and the equivalent circuit is the bridge for two measurements [29]. Therefore, the equivalent circuit helps differentiate and connect two measurements, playing a key role to clarify electron and proton process in the measuring system. Undoubtedly, this clarification can obtain complete and reliable circuit elements, helping reveal kinetic relationship for measuring system under successive temperature cycles. This will drive in situ characterization for in-plane conductance of membrane, thus inducing significant progress for materials and cells simultaneously.

## 2. EXPERIMENTAL SECTIONS

2.1 Materials and preparation

The ultrapure water with high electrical resistivity at 18Mohm cm (conductivity at 0.055μS/cm) was used in whole experiments. All glasswares were immersed into concentrated nitric acid to remove any residues for more than 24 hrs, and then they were cleaned in boiling water for more than six times.

The Nafion XL-100 membrane (DuPont, thickness of 27.5 μm) was used as standard sample, which can be prepared for fuel cell directly according to supplier's specification. Therefore, we just immersed the Nafion XL-100 membrane into ultrapure water before one-day measurement, helping attain its sufficient hydration state.

2.2 Measuring electrode and Fixture

To illustrate in-plane conductivity measurement, Fig.1 a exhibits measuring electrodes as Pt lines, with small diameter (Φ=0.5mm) towards exact measurement [13]. Figure 1b shows the measuring fixture with Nafion membrane, and the fixture was made by insulative PTFE materials with same top and bottom segments. In its middle section, the fixture has blank area to permit water transfer freely. On its opposite edges, the fixture has inner grooves to fasten Pt lines, with 1cm distance for adjacent lines.

2.3 Membrane Installation and operation

To install membrane in fixture (Figure 1b), the Nafion membrane was put on bottom segment of fixture, and its top surface was attached on four parallel Pt wires fastened by insulative tapes. Then, the top segment of fixture was covered on membrane, and the whole fixture was fastened by PTFE screws under torque of 1.5 N m . Subsequently, the whole fixture was immersed into ultrapure water bath covered by protection membrane. Finally, the fixture in ultrapure water bath was put into heating water bath, whose temperature was controlled to increase from 20 to 100°C in every day. Each temperature was kept at least 20 minutes for stability before measurement, and the whole measurement processes were carried out on 3 days successively.

2.4 The in-plane conductivity measurement for Nafion Membrane

The electrochemical measurement was carried on Solartron Energylab workstation,

with specific definition as $RE_1$, $RE_2$ corresponding to reference electrode (RE) and sensor electrode(SE), respectively. According to principle for 4-electrode measurement, the RE and SE were used to control potential as outside lines, while counter electrode (CE) and working electrode(WE) were used to reflect current as inside lines [25]. Figure 1c shows the 4-electrode composition as $RE_1$(RE)-CE-WE-$RE_2$(SE), with 1cm distance (d) for adjacent lines. Figure 1c also labelled membrane length (w= 5cm) and thickness (t= 27.5 μm), helping calculate in-plane conductivity.

We adopted cyclic voltammetry (CV) and EIS measurements under potential mode, corresponding to DC and AC measurement methods, respectively. On CV measurement, the potential was set as 0-1V range with scan rate of 20 mV/s. On EIS measurement, the frequency range was set as 1M Hz ~1m Hz, with amplitude of sinusoidal potential at 10 mV but without any DC potential deviation. With Zview software, the measured EIS dates were checked reliability under Kramers-Kronig transformation, and then they were fitted with equivalent circuit. Furthermore, the in-plane conductivity (σ, S cm$^{-1}$) of Nafion membrane was calculated as：

$$\sigma = \frac{d}{w \cdot t \cdot R}$$

Where, d is current line distance (1cm) between CE-WE, w and t are membrane length(5cm) and thickness (27.5 μm), and R is resistance (ohm) in EIS measurement.

3. RESULTS AND DISCUSSIONS

3.1 Basic features for CV and EIS measurement

To understand basic measurements, Fig. 2 illustrates typical CV and EIS results at 30

°C condition in 1st day. Fig. 2(a) shows that the CV under potential range of 0-1V can induce ~0.2-1.2 mA with a straight and reversible line, dominated by resistance at 1037.3ohm. This can be due to total resistance from electron and proton with successive transportation, thus dictating Ohm's law as Equation(1)

$$\eta_{ohmic}^{total} = \eta_{ohmic}^{electronic} + \eta_{ohmic}^{protons} = -i \ R^{interal} \qquad (1)$$

In the meanwhile, the membrane conductivity (K) is determined as Equation(2):

$$K = F^2 D_H^+ C_H^+ / RT \qquad (2)$$

where, F is Faraday's constant(96,487 C/mol), R is gas constant, 8.3143 J/(mol)(K). T is Kelvin temperature, $C_H^+$, $D_H^+$ are concentration and diffusivity of H+ species ($H^+ \cdot H_2O$), respectively.

Since the $C_H^+$, $D_H^+$ are dependent upon the water content and distribution in the membrane, the proton conductivity and resistance are constant[30]. Therefore, this simple CV measurement can provide resistance information, dictating Ohm's law in the heterogeneous system.

Fig. 2(b) further shows EIS measurement, whose Nyquist plot exhibits high frequency arc for capacitance (HF, 1000 K-2~6K Hz) at 1st quadrant, in contrast to low frequency (LF, 2~6 K- 0.3Hz) arc for inductance at 4th quadrant. From the view of electricity, the capacitance always help proton mobility but the inductance always inhibit proton mobility. Therefore, it is explicit that the intersection point at abscissa axis is the maximum of proton mobility, corresponding to membrane resistance at 1365.5ohm. These two opposing phenomena can be due to opposing proton behaviors on Nafion membrane. On the one hand, the proton mainly transport through water channels in membrane, dictating as proton mobility. On the other hand, the $HSO_3^-$ group in membrane has strong interaction with proton, thus inhibiting proton mobility to dictate mobility hysteresis behavior [22,31]. This clarification is very important to realize the proton mobility in Nafion membrane, thus having a general theoretical significance.

It should be stressed that this LF loop is a common feature in fuel cell operation, but this has been largely shielded due to complex cell operation [32,33]. Past studies found that the LF inductive is mainly related to slow kinetics for Nafion ionomer humidification[34], along with ionomer swelling/shrinking (hydration/dehydration) cycles in cathode catalyst layer [35--37]. Only recently, the high-frequency resistance measurement can capture the ohmic resistance accurately, corresponding to the frequency of zero-phase from 1 kHz -10 kHz [38]. This frequency range(1-10K Hz) strongly supporting 2-6K Hz in our approach, indicating it as a characterization frequency for humidified Nafion membrane. Therefore, our EIS measurement can obtain more information than that in CV measurement, penetrating into membrane structure insightfully.

3.2 Equivalent circuit buildup

Fig. 3a depicts that both DC and AC measurements can induce main response on contact interface (Pt electrodes and Nafion membrane) and Nafion membrane, inevitably dictating as Equivalent circuit. Fig. 3b proposed that the contact interfaces transfer heterogeneous charges, probably establishing parallel circuit for contact capacitor ($C_0$) and contact resistance ($R_0$) [39]. Fortunately, Fig.2b shows the high

frequency(HF) comes from original point if extrapolating high frequency linearly, indicating low contact resistance ($R_0$). This also means the higher frequency can drive quick charging/discharging process for contact capacitor ($C_0$), thus the contact capacitor can attain steady state to present low contact resistance ($R_0$). Therefore, Fig.3b only keeps contact resistance ($R_0$), without considering contact capacitor presence. Moreover, Fig.3a proposes the electric field (U) on membrane can establish three parallel processes in Fig.3b.

1$^{st}$, The electric field establishes charge reservoir, similar as capacitor to keep heterogeneous charges in membrane. Considering leakage current between measuring electrodes, Fig.3b used constant phase element (CPE) to represent capacitor.

2$^{nd}$, The electric field can drive active proton mobility in water channels, presenting proton resistance ($R_1$).

3$^{rd}$, The electric field cannot drive inactive proton mobility in $HSO_3^-$ group. Since this dictates proton mobility hysteresis in LF region, Fig.3b uses series path of $R_2$-$L_2$ to represent inactive resistance and inductance.

Therefore, Fig. 3b proposes equivalent circuit with explicit physical meanings for all elements, helping check various measurements.

3.3 Validation of the Equivalent circuit

After successive measurements, we checked all EIS dates reliability and then fitted them with equivalent circuit. Fig 4 only exhibits the measured and fitted Nyquist spectrums in 1$^{st}$ day, illustrating their high fitting manners. Table 1-3 further lists all fitted parameters under successive temperature cycles in 3 days, showing fitting with total Chi-square errors less than 0.4253. These results support the equivalent circuit is effective to reflect EIS measurement, helping reveal measuring system completely and dynamically.

As a typical DC approach, the CV measurement experiences open path for capacitance and short path for inductance in the equivalent circuit. Therefore, the steady CV measurement only makes total resistance response in equivalent circuit, dictating as $R_0+R_1R_2/(R_1+R_2)$. Fig.5 compares the measured resistance in CV measurement and the fitted resistance of $R_0+ R_1R_2/(R_1+R_2)$ in EIS measurement (Table 1-3), indicating their high similarity even under successive temperature cycles. This similarity not only makes inter-corroboration for two measurement, but also validates general effectiveness for equivalent circuit.

With its bridge role, the equivalent circuit should differentiate CV and EIS response. As shown before, the steady CV measurement makes total resistance response as $R_0+R_1R_2/(R_1+R_2)$. From equivalent circuit, the parallel resistance for $R_1$ and $R_2$ is smaller than individual resistance, leading to $R_0+ R_1R_2/(R_1+R_2)<R_0+R_1$. Noticeably, Table (1-3) show all $R_0$ values are smaller than other resistances, suggesting $R_0+R_1R_2/(R_1+R_2)<=R_1$. This general relationship can be found in Fig. 2, where the $R_0+R_1R_2/(R_1+R_2)$ resistance at 1037.3 ohm in CV measurement is smaller than $R_1$ resistance at 1365.5ohm in EIS measurement.

The above clarification supports that the CV measurement has total resistance response, driving both inactive and active proton transportation in equivalent circuit. Furthermore, the Ohm's law can apply for measuring system with heterogeneous interface, extending the traditional Ohm's law range in metal system with good electron conductivity. Moreover, the above clarification makes explicit that the EIS measurement mainly use multi frequency to reveal various electric responses, helping obtain equivalent circuit directly. Consequently, both CV and EIS measurement can support reliable equivalent circuit with inter-corroboration, helping reflect measuring system completely and dynamically. This general electrochemistry approach helps reveal kinetic physical-chemistry process in measuring system, thus providing In Situ characterization technique insightfully.

3.4 The circuit elements analysis for kinetic relationship.

3.4.1 The contact resistance

Based on equivalent circuit validity, the circuit element can disclose measuring system under dynamical operation. Fig 6 demonstrates the contact resistance ($R_0$) has small values of -20-40 ohm, indicting tight contact interface (Pt wire/membrane) under successive temperature cycles. Fig 6 further shows the $R_0$ fluctuated largely below 80 ˚C but kept stable after 80 ˚C, indicating challenge interface change.

Generally, the Nafion membrane has glass transition temperatures (Tg) at 80 ~ 85 °C, where the main- and side-chains of membrane occur long-range mobility. This mobility may last at 110 °C, dictating as α-relaxation process[31]. Furthermore, the Nafion membrane has high swelling ratio above 80°C, and its dimension usually changes in the through-plane direction. This helps tight membrane towards measuring electrodes, thus keeping steady contact resistance ($R_0$) [24]. In contrast, past studies found in the cooling run, the membrane may occur **β**-relaxation process towards main-chain motions at 35 °C, and this **β**-relaxation process should relief stress at 70–75 °C in the following heating run [40]. Therefore, in the temperature cycles below 80 °C, the membrane usually occur re-organization, thus its unstable swelling ratio can dictate significant fluctuation for contact resistance ($R_0$). Consequently, our kinetic relationship for contact resistance provides dynamical characterization for hard electrode-soft membrane, relating with advanced characterizations strongly.

3.4.2 The CPE on membrane

Under successive EIS measurement, the electric field on membrane can be revealed by constant phase element (CPE) with resistance $Z_Q = \frac{1}{Y_0}(j\omega)^{-n}$, where $Y_0$ is reflection signal and "n" can index special process. Table 1-3 show all CPE-n lie in 0.89-0.99, indicating rough or porous membrane to establish ideal capacitor (n=1) [41]. This further suggested that the electric field establishes on rough surface and porous structure in membrane, promising in situ characterization completely and dynamically.

3.4.3 The in-plane conductivity for Nafion membrane

Fig. 7 (a-c) compare the in-plane proton resistance measured at the characteristic frequency (2-6K Hz) and fitted resistance ($R_1$) in equivalent circuit, dictating their high similarity under successive temperature cycles. This strongly supports the characteristic frequency (2-6K Hz) is a simple tool to obtain Nafion membrane resistance, helping fuel cell evaluation directly.

Fig.7(a-c) further exhibits the increasing temperature can decrease proton resistance while increase proton conductivity monotonically. This kinetic relationship can be realized in two aspects, Nafion membrane and water media condition. On the one hand, the membrane experience slight structure change below 80°C where **β**-relaxation process help membrane re-organization. When temperature is higher than 80°C, the membrane began to experience glass transition process, thus the α-relaxation process can increase internal volume and outer dimension, along with reduction of ionomer density in membrane remarkably[24,40,42-44]. On the other hand, the increasing temperature can gradually increase water activity in water media, shifting water diffusion towards convection in operation condition.

Based on the above clarification, it can be determined that at low temperature, the water media mainly adopt Fick diffusion into membrane, whose surface and thickness can strongly inhibit water diffusion towards bulk polymer[26,45,46]. In the meanwhile, the membrane surface is enrich of sulfonic acid ionic clusters, leading to strong surface conductivity [23,26, 46-48].

With the temperature increasing, the water media can shift diffusion towards convection, and the membrane can decrease its pore fluid viscosity gradually. This can accelerate water transport towards bulk polymer, making conductive pathways and connectivity network smoothly [49]. Simultaneously, the increasing temperature can reduce activation barrier for proton transportation between hydrophilic clusters of membrane, driving bulk conductivity [3]. This case is more serious when attaining glass temperature (80°C), where membrane can enlarge internal volume to permit more water transportation significantly. Due to above reasons, the increasing temperature can decrease proton resistance while increase proton conductivity for Nafion membrane monotonically in water media.

To provide high-standard comparison, we selected critical 80 ˚C condition as baseline, dictating in-plane conductivities at 0.12, 0.115, 0.1 S cm$^{-1}$ under successive 3 days. These values are highly consistent with 0.12 S cm$^{-1}$ in liquid water media at 80 ˚C [13], driving measurement towards standardization. Fig 7 further shows the in-plane conductivities in 20 -100°C lie in 0.04-0.15 S cm$^{-1}$ in 1$^{st}$ day, 0.06-0.13 S cm$^{-1}$ in 2$^{nd}$ day, 0,06-0.118 S cm$^{-1}$ in 3$^{rd}$ day. This fluctuation can be due to that the successive temperature cycles can drive bulk polymer loss and volume change, thus adjusting conduction channels slightly [3,24,,50,51].

3.4.4 The LF loop

Under successive temperature cycles, Fig 8 exhibits the $R_2$ resistance and $L_2$ inductance fluctuation significantly, coming from low frequency (LF) loop in EIS measurement. From electrical opinion, the $L_2$ inductance always induces counter electromotive force under electric fluctuation, helping keep steady measurement process. In this case, only the $R_2$ resistance can dictate inactive proton mobility with strong interaction of sulfuric groups [3233], in contrast to $R_1$ resistance as active proton mobility in water channels. Furthermore, the large fluctuation of $R_2$ in the range of 2000-8000ohm (Fig 8) indicates that the inactive proton shifts towards active proton, thus maintaining proton mobility effectively. Therefore, this shift helps characterize water role in membrane, usually dictating conductivity hysteresis in LF loop [12,52].

It should be stressed that this low frequency(LF) loop in EIS measurement is very similar to DC measurement without any frequency, making it as central issue in electrochemistry. Furthermore, the LF inductance always means some inductance can induce counter electromotive force, suggesting it is an important sign for stable system [53,54]. As shown before, the LF inductive is mainly related to Nafion ionomer humidification [34] and swelling/shrinking (hydration/dehydration) cycles, helping study the challenge tri-phase boundary in cell operation [35-37]. Therefore, our LF loop on Nafion membrane provides core material support for active and stable fuel cell, also helping clarify water role in cell operation.

Our LF inductance has great reference for secondary battery, in which the lead/acid battery[55] and lithium-ion battery [56] reported their LF inductive on electrode/electrolyte interfaces. Furthermore, our LF inductance has great reference for various kinds of materials. For metal materials, the high-Entropy alloy and 304 stainless steel usually experience surface passivation along with pitting from aggressive $Cl^-$ in NaCl solution. These pitting- passivation phenomena can be identified in LF loop, supporting stable system under activation process [57]. For semiconductor materials, the p-type silicon in etching process can form submonolayer oxide on surface, whose dynamical change can be monitored by LF loop [58]. For composite materials preparation, the montmorillonite sheets under electrolyte effect can be identified in the LF loop [59]. All these studies suggest the LF inductance is an important characterization sign for stable interface/surface under dynamic interaction, thus their electric phenomena can be usually observed in cells and secondary batteries.

Conclusion

It is very important to measure in-plane conductivity of Nafion membrane for fuel cell, but this target is generally inhibited by measuring system with heterogeneous interface and immature electrochemical measurements. This paper used water media to establish stable system with metal electrode and Nafion membrane, and further representing system as equivalent circuit. Our proposed equivalent circuit is validated by both CV measurement and EIS measurement, whose connection and difference can be clarified by the equivalent circuit. Therefore, the equivalent circuit provides in situ characterization tool to reflect system completely, and the circuit elements can penetrate into special physical process kinetically. For system under successive temperature cycles, the circuit element disclose the contact resistance changes dynamically due to relaxation process in membrane. The circuit element also determines the in-plane conductivity is surface conductivity, which shifts towards bulk conductivity with the temperature increasing, strongly relating to membrane structure and water media condition. Inspiringly, the elements found the inactive proton shifts towards active proton, helping maintain proton mobility in Nafion membrane. We also clarified this shift presents in low frequency (LF) inductance, which is an important sign to study stable and active operation for fuel cell, secondary battery and various kinds of materials. Consequently, our equivalent circuit make core breakthrough to measure in-plane conductivity of Nafion membrane, driving various measurement methods as in situ characterization tool to realize measuring system. This breakthrough can induce significant paces for multi-disciplines, thus having great significance and broad impact for future studies. .

*Acknowledgments*. This work is supported by the National Key Research and Development Program of China (2019YFB1504503), and National Natural Science Foundation of China (50873050, 51273103,51573084,21776158).

MATCHEMPHYS-D-22-06052-1st-revision

...

Table 1 The fitted parameters from equivalent circuit for EIS in 1$^{st}$ day

| parameters | $R_0$ (ohm) | CPE-$Y_0$ (ohm$^{-1}$ cm$^{-2}$ s$^{-n}$) | CPE-n | $R_1$ (ohm) | $R_2$ (ohm) | $L_2$ (Henry) | Chi-square |
|---|---|---|---|---|---|---|---|
| 20 °C | -12.73 | 6.7*10$^{-10}$ | 0.9738 | 1402 | 18257 | 22.52 | 0.4253 |
| 30 °C | 0.8548 | 4.99*10$^{-10}$ | 0.9965 | 1342 | 4407 | 68.82 | 0.0322 |
| 40 °C | 5.029 | 6.05*10$^{-10}$ | 0.9825 | 1153 | 6656 | 1676 | 0.0202 |
| 50 °C | 5.742 | 6.71*10$^{-10}$ | 0.9759 | 966 | 6275 | 1024 | 0.0283 |
| 60 °C | 13.87 | 6.23*10$^{-10}$ | 0.9823 | 800 | 8514 | 2292 | 0.0863 |
| 70 °C | 20.24 | 7.19*10$^{-9}$ | 0.9779 | 662 | 6323 | 647 | 0.0399 |
| 80 °C | 15.58 | 1.72*10$^{-9}$ | 0.9253 | 572 | 3454 | 364.2 | 0.0374 |
| 90 °C | 11.99 | 2.23*10$^{-9}$ | 0.9123 | 514 | 3006 | 158 | 0.0267 |
| 100 °C | 15.18 | 1.71*10$^{-9}$ | 0.9329 | 455 | 1645 | 670 | 0.0520 |

Table 2 The fitted parameters from equivalent circuit for EIS in 2nd day

| parameters | $R_0$ (ohm) | $CPE\text{-}Y_0$ (ohm$^{-1}$ cm$^{-2}$ s$^{-n}$) | CPE-n | $R_1$ (ohm) | $R_2$ (ohm) | $L_2$ (Henry) | Chi-square |
|---|---|---|---|---|---|---|---|
| 20 °C | 4.209 | 1.95*10$^{-9}$ | 0.9209 | 1224 | 3660 | 3508 | 0.009 |
| 30 °C | -15.9 | 2.94*10$^{-9}$ | 0.8903 | 1202 | 8108 | 8180 | 0.056 |
| 40 °C | 5.111 | 2.24*10$^{-9}$ | 0.9133 | 979 | 7579 | 3723 | 0.039 |
| 50 °C | 19.01 | 1.91*10$^{-9}$ | 0.9250 | 844 | 4621 | 3014 | 0.007 |
| 60 °C | 14.03 | 1.95*10$^{-9}$ | 0.9223 | 752 | 4130 | 4110 | 0.010 |
| 70 °C | -4.96 | 2.64*10$^{-9}$ | 0.8957 | 707 | 4175 | 2036 | 0.011 |
| 80 °C | 15.58 | 1.56*10$^{-9}$ | 0.9404 | 642 | 2318 | 709 | 0.015 |
| 90 °C | 12.51 | 1.78*10$^{-9}$ | 0.9304 | 596 | 2582 | 287 | 0.026 |
| 100 °C | 14.90 | 1.96*10$^{-9}$ | 0.9226 | 538 | 1625 | 698 | 0.038 |

Table 3 The fitted parameters from equivalent circuit for EIS in 3rd day

| parameters | $R_0$ (ohm) | CPE-$Y_0$ (ohm$^{-1}$ cm$^{-2}$ s$^{-n}$) | CPE-n | $R_1$ (ohm) | $R_2$ (ohm) | $L_2$ (Henry) | Chi-square |
|---|---|---|---|---|---|---|---|
| 20 °C | 18.1 | 2.12*10$^{-9}$ | 0.9268 | 1419 | 5922 | 3052 | 0.0316 |
| 30 °C | 31.08 | 2.00*10$^{-9}$ | 0.9345 | 1356 | 4786 | 1739 | 0.0250 |
| 40 °C | 20.96 | 1.90*10$^{-9}$ | 0.9361 | 1205 | 4687 | 793 | 0.0224 |
| 50 °C | 32.36 | 1.91*10$^{-9}$ | 0.9378 | 1025 | 4219 | 946 | 0.0205 |
| 60 °C | 20.50 | 3.05*10$^{-9}$ | 0.8979 | 905 | 5288 | 638 | 0.0210 |
| 70 °C | 18.46 | 2.13*10$^{-9}$ | 0.9143 | 777 | 4401 | 715 | 0.0123 |
| 80 °C | 6.876 | 2.47*10$^{-9}$ | 0.8977 | 712 | 5224 | 674 | 0.0100 |
| 90 °C | 11.84 | 2.30*10$^{-9}$ | 0.9028 | 641 | 3796 | 1199 | 0.0100 |
| 100 °C | 5.52 | 3.22*10$^{-9}$ | 0.8816 | 611 | 4598 | 856 | 0.0074 |

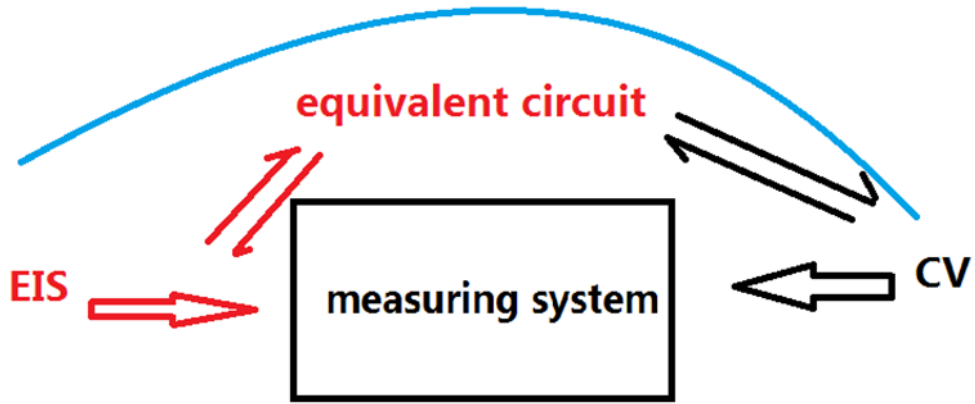

Scheme.  The proposed study scheme in this work

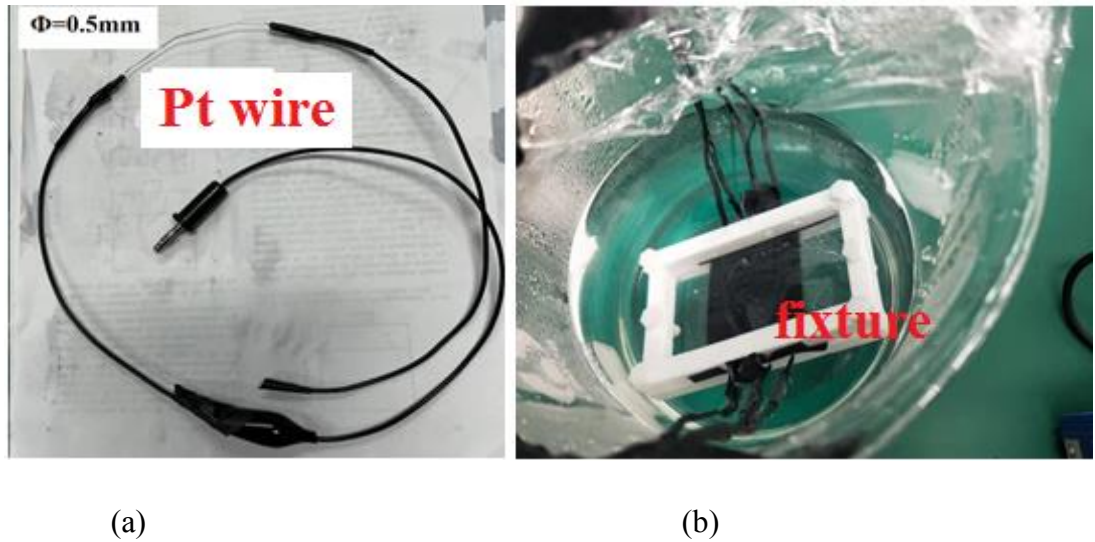

(a)   (b)

(c)

Fig.1 The illustration for in-plane conductivity measurement; (a) Pt wire electrode, (b)The fixture with Nafion membrane, (c) The electrode and Nafion membrane composition, where d is current line distance (1cm), w and t are membrane length(5cm) and thickness (27.5 μm)

(J.W. GUO et al.)

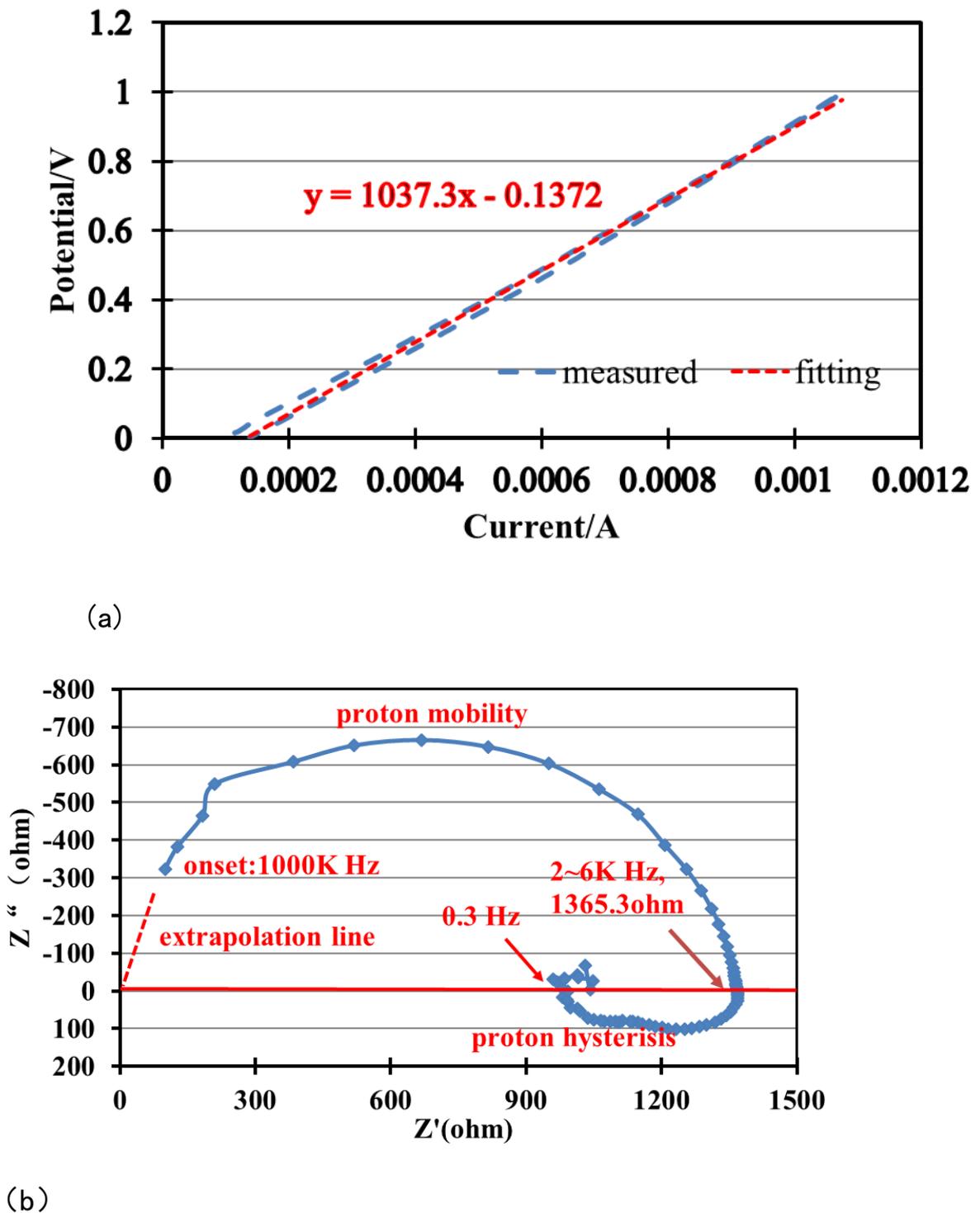

(a)

(b)

Fig.2 The in-plane measurement for Nafion membrane at 30 ˚C condition, 1st day: (a) CV plot, (b) Nyquist plot in EIS measurement.

(J.W. GUO et al.)

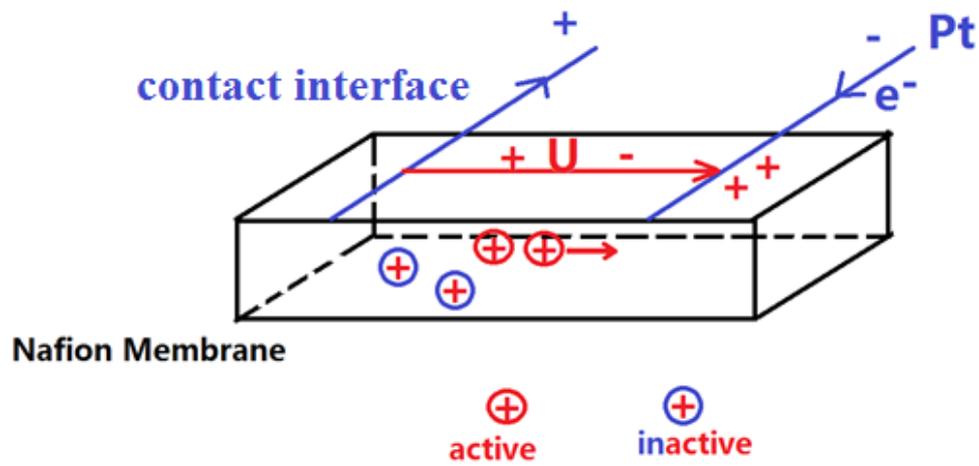

(a)

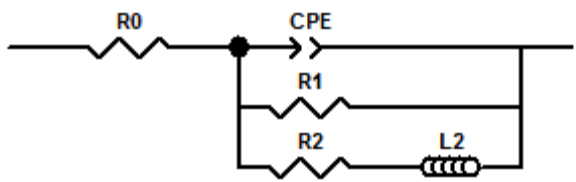

(b)

Fig.3 (a)The measurement under both DC and AC approaches.

(b)The proposed equivalent circuit

(J.W. GUO et al.)

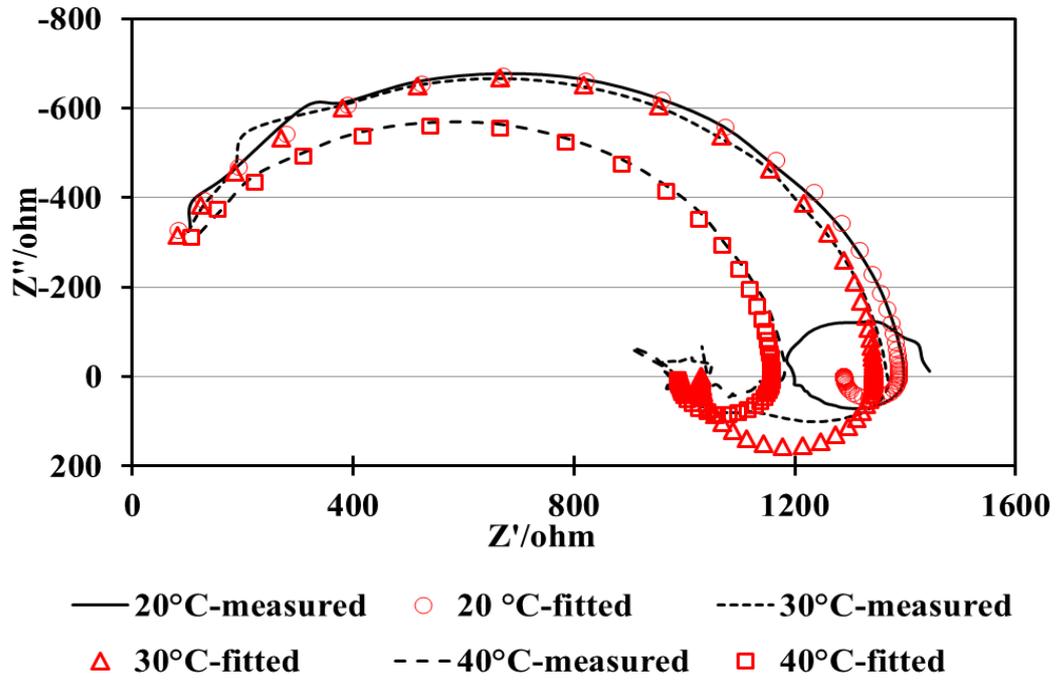

(a)

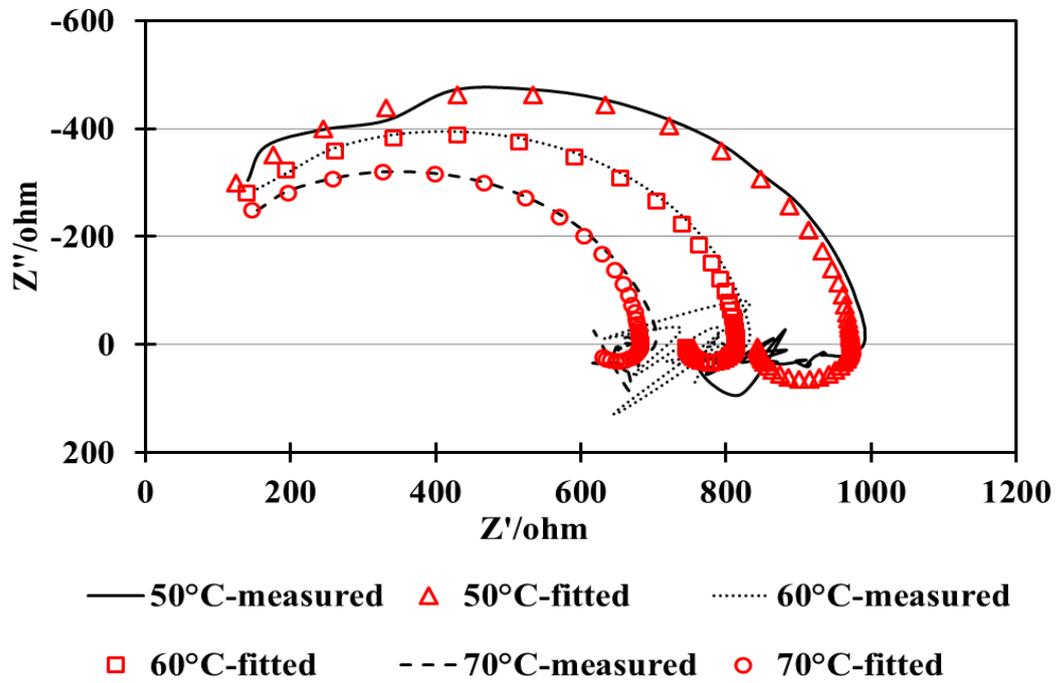

(b)

(J.W. GUO et al.)

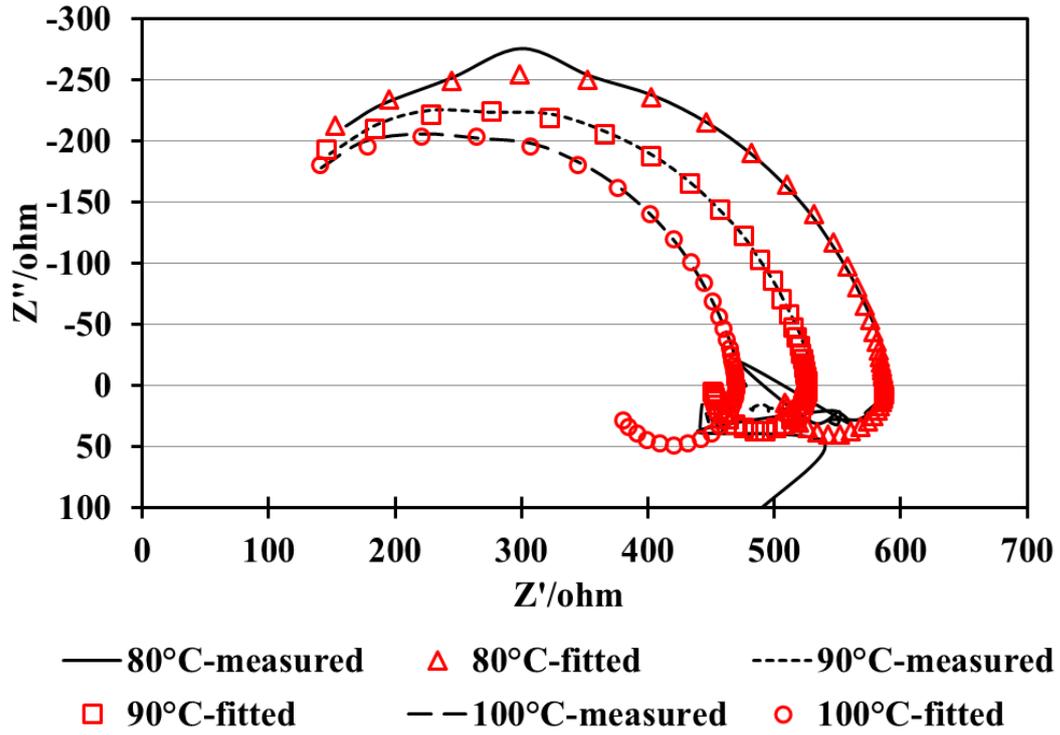

(c)

Fig.4 The temperature effect on measured and fitted Nyquist spectrum in 1ˢᵗ day. (a) 20-40 ˚C; (b)50-70 ˚C;(c) 80-100 ˚C.

(J.W. GUO et al.)

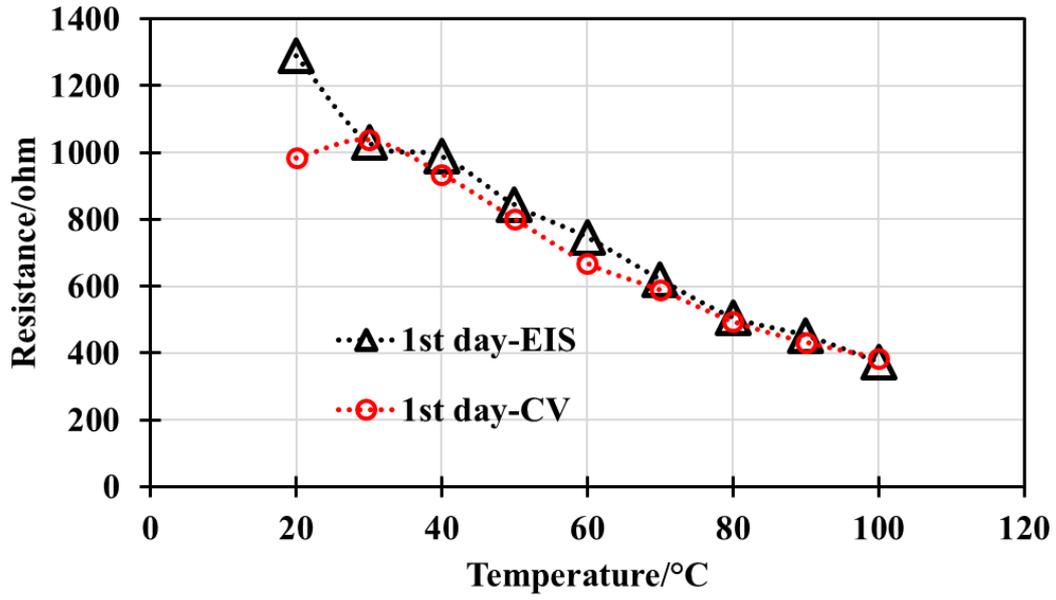

(a)

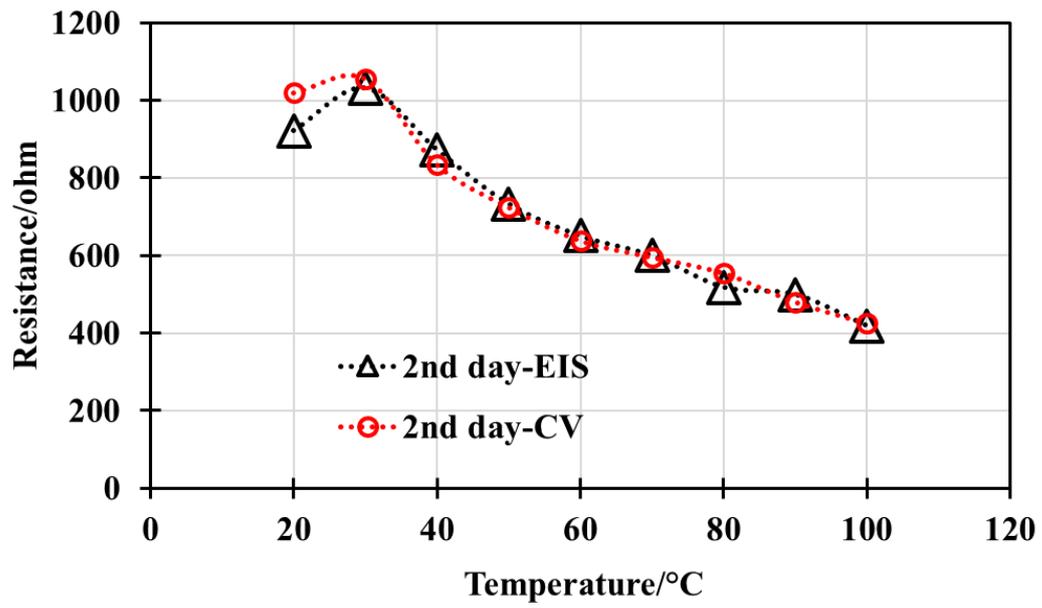

(b)

(J.W. GUO et al.)

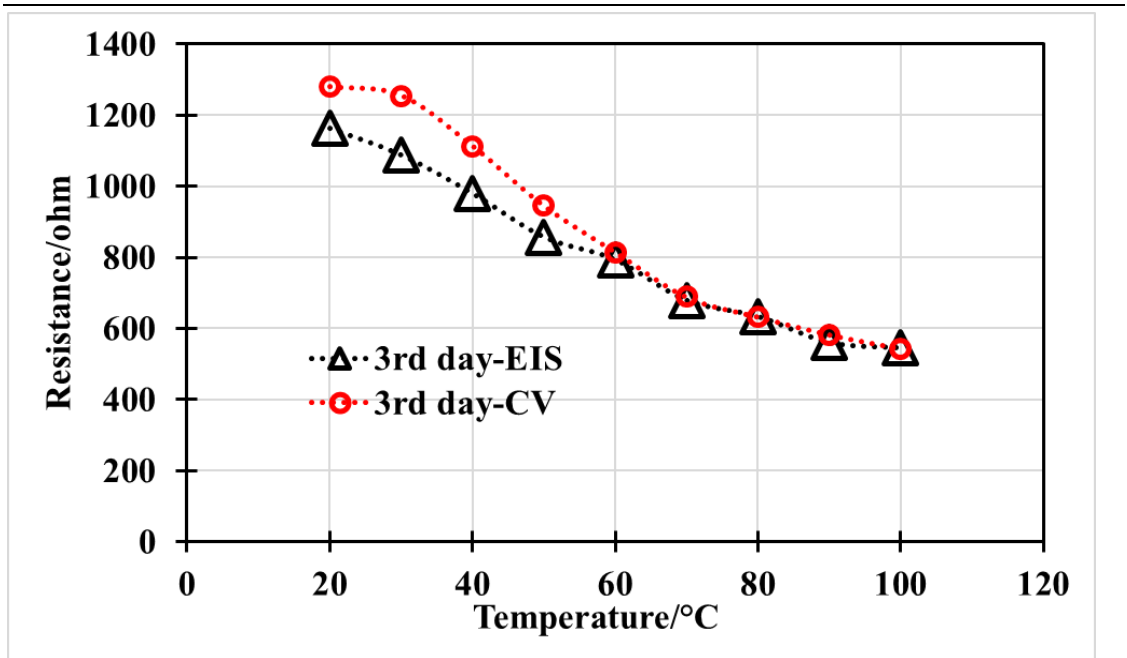

(c)

Fig.5 The comparison for the measured resistance in CV measurement and the fitted resistance of $R_0 + R_1R_2/(R_1+R_2)$ from EIS measurement under successive temperature cycles. (a) 1st day; (b) 2nd day; (c) 3rd day.

(J.W. GUO et al.)

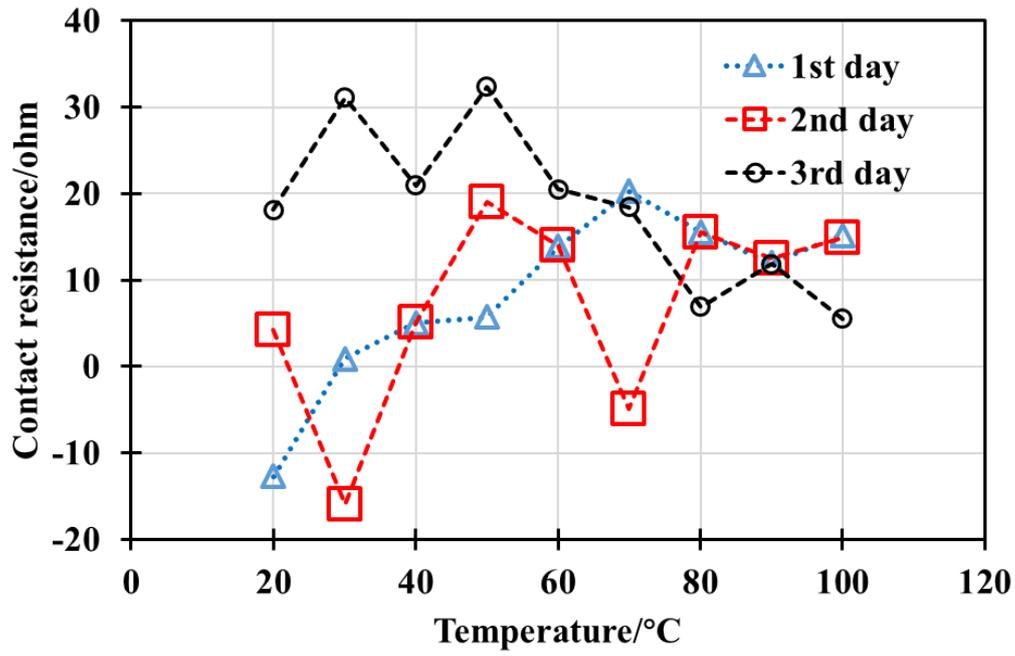

Fig. 6 The contact resistance($R_0$) under the successive temperature cycles

(J.W. GUO et al.)

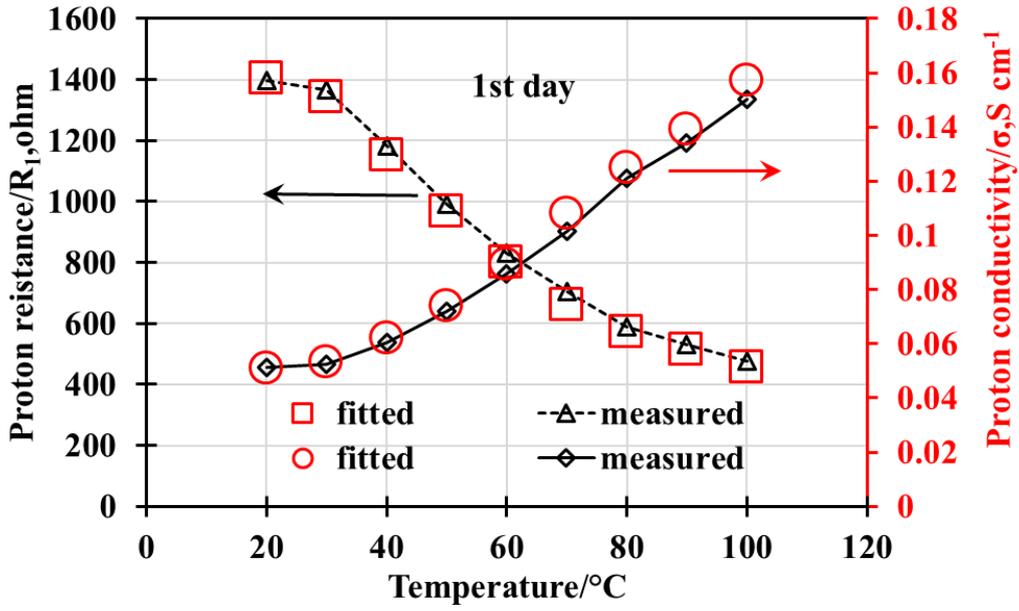

(a)

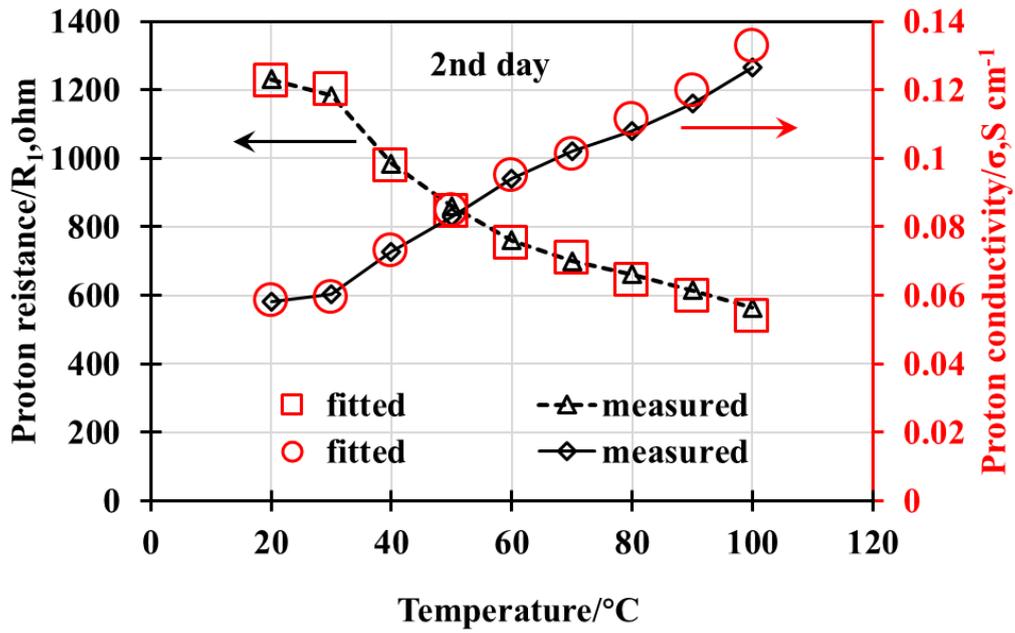

(b)

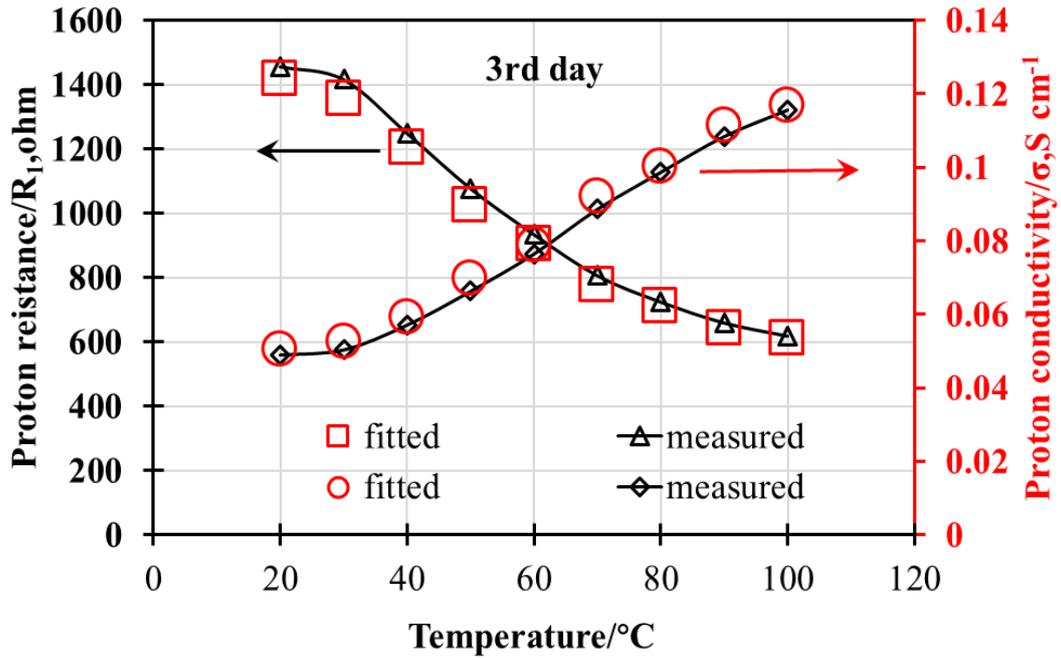

(c)

Fig.7 The measured (2-6K Hz) and fitted proton resistances($R_1$, ohm), proton conductivity ($\sigma$, S cm$^{-1}$) under successive temperature cycles. (a) 1st day. (b) 2nd day; (c) 3rd day

(J.W. GUO et al.)

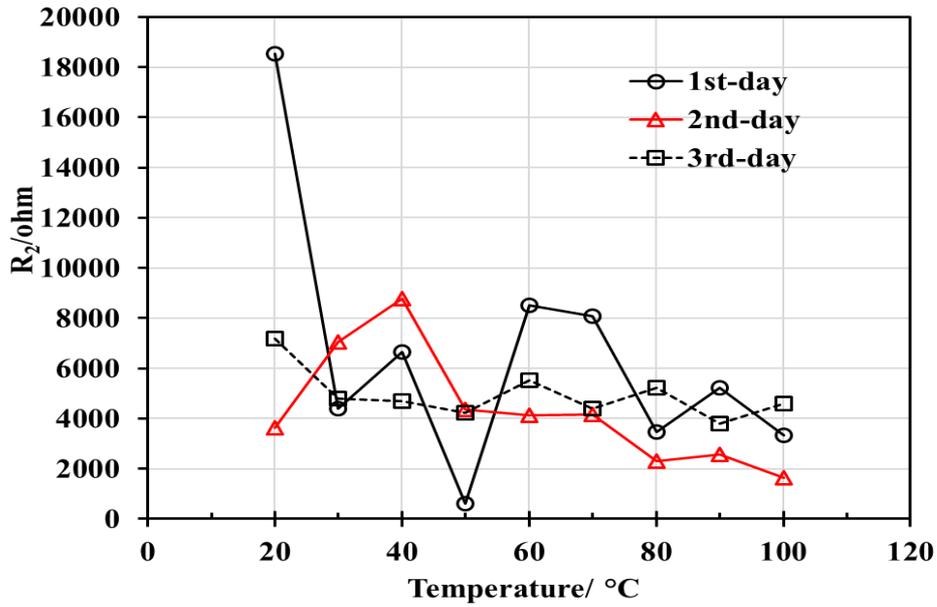

(a)

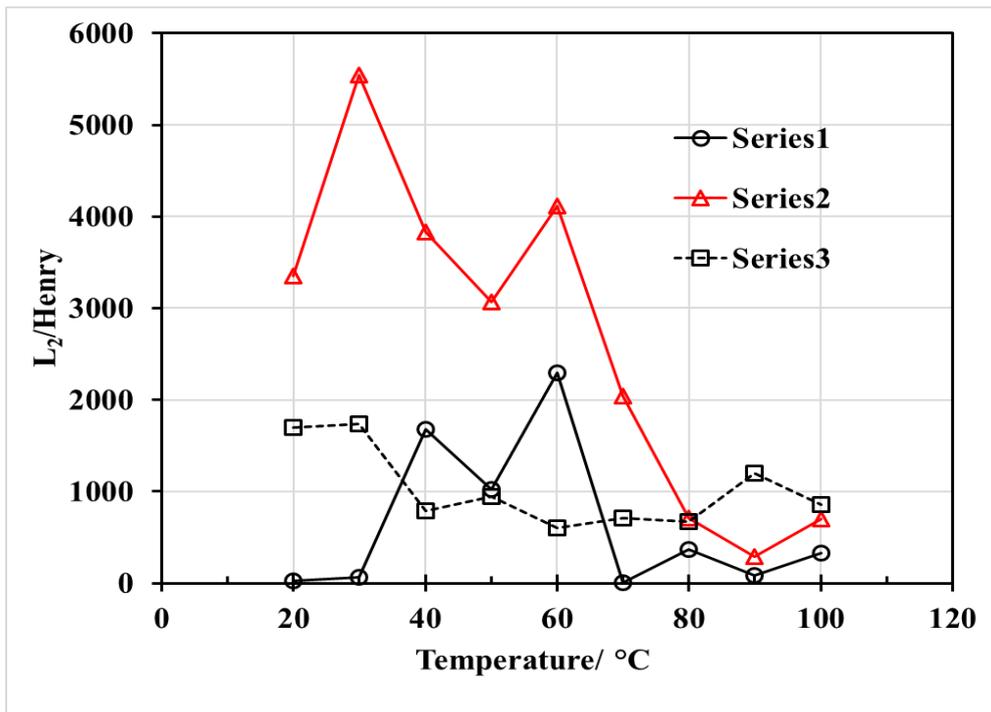

(b)

Fig.8　The effect of successive temperature cycles on the water hysteresis in Nafion membrane. (a)The resistances($R_2$). (b) the inductance($L_2$)

(J.W. GUO et al.)